\documentclass[pre,showpacs,onecolumn,amsfonts,amsmath,amssymb,floatfix]{revtex4}
\usepackage{amssymb}
\usepackage{graphicx}% Include figure files
\usepackage{dcolumn}% Align table columns on decimal point
\usepackage{bm}% bold math
\usepackage{latexsym}
\usepackage{float}
\usepackage{supertabular}
\usepackage{longtable}
\usepackage{color}
\usepackage{float}
\begin{document}
\title{Community Structure of the Physical Review Citation Network}
\author{P.~Chen}
\author{S.~Redner}
\affiliation{Center for Polymer Studies, and Department of Physics, Boston
  University, Boston, MA, 02215}

\begin{abstract}

  We investigate the community structure of physics subfields in the citation
  network of all Physical Review publications between 1893 and August 2007.
  We focus on well-cited publications (those receiving more than 100
  citations), and apply modularity maximization to uncover major communities
  that correspond to clearly-identifiable subfields of physics.  While most
  of the links between communities connect those with obvious intellectual
  overlap, there sometimes exist unexpected connections between disparate
  fields due to the development of a widely-applicable theoretical technique
  or by cross fertilization between theory and experiment.  We also examine
  communities decade by decade and also uncover a small number of significant
  links between communities that are widely separated in time.

\end{abstract}
\maketitle

\section{Introduction}

In this work, we study the community structure of citations within the
Physical Review (PR) family of journals from its inception in 1893 until
August 2007.  The journal consisted of just the Physical Review through 1969
and then split into 4 branches in 1970: Physical Review A (PRA): atomic,
molecular, optical physics; Physical Review B (PRB): solid-state,
condensed-matter physics; Physical Review C (PRC) nuclear physics; Physical
Review D (PRD): particle physics.  In 1990, there was yet another split of
Physical Review E (PRE): statistical physics from PRA.  The journal also
includes a letters section, Physical Review Letters (PRL), that was
introduced in 1958, a review section, Reviews of Modern Physics (RMP), that
was introduced in 1929, and two recent special topics journals.  For PR
publications, we ask: can articles be naturally grouped into distinct
subfields, with a high density of citations among papers within a given
subfield and sparser citations across subfields?  By the very nature of
physics research and also as revealed by the data, this partitioning into
subfields is self-evident.  While anecdotal information exists about the
identity and evolution of some of the more prominent subfields of physics
\cite{anec}, here we determine their quantitative properties, such as their
size, time history, and citation impact.  A related work recently studied the
evolution of scientific fields through the PACS (Physics and Astronomy
Classification Scheme) numbers of each article~\cite{CS1}.

There are a variety of compelling reasons for studying the community
structure in complex networks.  In social networks, the partitioning of
acquaintances into communities represents a basic fact about human
interactions~\cite{social}.  In metabolic networks, community structure may
help identify basic reaction modules \cite{GN}.  In the web, community
structure reveals connections to web pages on related topics \cite{WWW}.  For
the citation network, its underlying community structure may help us
understand both the obvious and the more subtle interrelations between
subfields, as well as the growth and the ebb of subfields.

To identify communities within networks, a variety of methods have been
developed, such as the Kernighan-Lin algorithm~\cite{KL}, spectral
partitioning~\cite{GP1,GP2}, and hierarchical clustering~\cite{HC1,social}.
While these methods sometimes work reasonably well, they can fail to identify
communities when applied to networks that lie outside of the domain of their
immediate application~\cite{Newman2}.  More recent work has led to the
formulation of new and powerful methods to detect communities in complex
networks, both with undirected~\cite{Newman4,CS1,CS2,CS3,CS4,CS5} and
directed~\cite{Newman4} links.  A systematic review of these developments is
given in~\cite{CS4}.  One particularly useful approach exploits the concept
of modularity~\cite{Newman}.  Compared to the earlier community detection
methods, the use of this metric to identify communities requires no extra
knowledge beyond the network structure itself, involves no subjective
judgments, and can be applied to any type of network.

In the next section, we outline the modularity maximization approach that we
use to resolve community structure.  We test the robustness of this method by
using an alternative bottom-up network community partitioning
algorithm~\cite{Vincent}.  We also check the significance of our results by
applying community detection to a randomized version of the citation network.
In Sec.~\ref{sec:PR}, we apply this algorithm to the Physical Review citation
network to resolve its major communities and the connections between them.
In Sec.~\ref{sec:Single}, we study the structure of individual communities.
In Sec.~\ref{sec:TV}, we partition PR publications into eight decadal sets by
year of publication: 1927--1936, 1937--1946, $\ldots$, 1997--2006 to study
the time evolution of physics fields.  Finally, we summarize in
Sec.~\ref{sec:Summary}

\section{Community Detection by Modularity Maximization}
\label{sec:MM}

For detecting communities within networks, we want to determine sets of
vertices that are more strongly connected to each other but less connected to
the rest of the network.  For this purpose, we use the modularity
$Q$~\cite{Newman3}, that is defined by
\begin{equation}
\label{modularity}
Q = \frac{1}{2L} \sum_{i,j} \Big[A_{i,j}-\frac{k_i k_j}{2L}\Big]\,\,\delta (i,j).
\end{equation}
Here $A_{i,j}$ is the $ij^{\rm th}$ element of the adjacency matrix
($A_{i,j}=1$ if a link exists between $i$ and $j$ and ($A_{i,j}=0$
otherwise), $L$ is the total number of network links, $k_i$ is the degree of
node $i$, and $\delta (i,j)$ equals $1$ if $i$ and $j$ belong to the same
group, and equals $0$ otherwise.  The modularity $Q$ gives the difference
between the number of links between groups in the actual network and the
expected number of links between these same groups in an equivalent random
network with the same link density.  A modularity $Q=0$ corresponds to a
random network, in which two nodes are connected with probability that is
proportional to their respective degrees.  Empirical data indicates that a
modularity value $Q\agt 0.3$ is indicative of true community
structure~\cite{M_example1,M_example2}, and the largest modularity that has
been observed in real-world examples is $0.7$~\cite{Newman}.  Thus we use
modularity maximization as the criterion to divide a network into
communities.

For large networks it is computationally impractical to maximize the
modularity over all possible partitions of the network and one must resort to
approximate methods~\cite{M_method}.  Here we apply the eigenvector approach
of Newman~\cite{Newman}.  Suppose that the network contains $N$ nodes and $L$
links.  We first focus on dividing the network into two communities, and then
generalize to an arbitrary number of groups.  Denote the two communities as 1
and 2, and let $s_i =1$ if node $i$ belongs to group $1$ and $s_i =-1$ if $i$
belongs to $2$.  Eq.~\eqref{modularity} can be rewritten as
\begin{eqnarray}
\label{Q-matrix}
Q & = & \frac{1}{2L} \sum_{i,j} \Big[A_{i,j}-\frac{k_i k_j}{2L}\Big]\,
\frac{1}{2}(1+s_{i}s_{j}) \nonumber\\
  & = & \frac{1}{4L} \sum_{i,j} \Big[A_{i,j}-\frac{k_i k_j}{2L}\Big]s_i s_j
   \equiv  \frac{1}{4L}\,\,{\bf s^{T}Bs}\,.
\end{eqnarray}
In going to the second line, we use $\sum_{i,j}A_{i,j} =2L$, so that
$\sum_{i,j} \Big[A_{i,j}-\frac{k_i k_j}{2L}\Big] =0$.  In the last line of
Eq.~\eqref{Q-matrix}, ${\bf s}$ is the vector whose elements are the $s_i$,
and ${\bf B}$ is the symmetric modularity matrix with elements
\begin{equation*}
B_{i,j}=A_{i,j}-\frac{k_i k_j}{2L}~.
\end{equation*}

The sum of each row and column of ${\bf B}$ equals zero, so that
$(1,1,1,\ldots,1)$ is necessarily an eigenvector of this matrix with zero
eigenvalue.  Let ${\bf u}_i$ be the complete orthonormal set of eigenvectors
of ${\bf B}$.  We can then write ${\bf s} = \sum_i a_i {\bf u}_i$, with $a_i
= {\bf u}^T_i \cdot {\bf s}$, so that the modularity becomes
\begin{equation}
\label{m2}
Q = \sum_i a_i {\bf u}^T_i\,\, {\bf B}\,\, \sum_j a_j {\bf u}_j =
\sum_i ({\bf u}^T_i \cdot {\bf s})^2 \beta _i ~.
\end{equation}
Here $\beta _i$ is the (ordered) eigenvalue of ${\bf B}$ that corresponds to
the eigenvector ${\bf u}_i$,with $\beta _1 \geq \beta _2 \geq \ldots \geq
\beta _n$.  To maximize the modularity, Eq.~\eqref{m2} suggests that we
choose ${\bf s}$ to be parallel to ${\bf u}_1$.

For the citation network, we need to extend the above approach to directed
networks, in which all links are directed by virtue of publications being
able to cite in the past.  For this purpose, we use a generalization, also
developed by Newman~\cite{Newman_d}, in which the modularity is now given by
\begin{eqnarray}
\label{Q_d}
Q   =  \frac{1}{2L} \sum_{i,j} \Big[A_{i,j}-\frac{k^{\rm in}_i
  k^{\rm out}_j}{L}\Big]\,\,(1+s_i s_j)
 \equiv  \frac{1}{2L} \sum_{i,j} s_i B^{(d)}_{ij} s_j ~,
%=  \frac{1}{2L} {\bf  s^{T}B^{(d)}s}.
%    & = & \frac{1}{4m} {\bf s^{T}(B+B^T)s}
\end{eqnarray}
where the elements of the directed modularity matrix ${\bf B}^{(d)}$ are:
\begin{equation*}
\label{B_d}
B^{(d)}_{ij} = A_{ij}-\frac{k^{\rm in}_i k^{\rm out}_j}{L}
\end{equation*}
We thus maximize the modularity in Eq.~\eqref{m2} using the eigenvalues and
eigenvectors of the directed modularity matrix.

Since the value of each $s_i$ can only be $1$ or $-1$ by definition, ${\bf
  s}$ cannot be parallel to ${\bf u}_1$ except for special conditions.  A
natural alternative is to choose the sign of $s_i$ according to the sign of
its corresponding element in ${\bf u}_1$.  For example, if the elements
$n_1,n_2,\ldots, n_j$ of ${\bf u}_1$ are negative while all rest are
positive, we choose $s_{n_1}=s_{n_2}=\ldots s_{n_j}=-1$ and all other $s_i$
equal to $+1$.  This partitioning then also determines the sizes of the two
groups.  As long as the largest eigenvalue is positive, there is a meaningful
division of the network into subgroups.  However, when the largest eigenvalue
is zero, which corresponds to the eigenvector $(1,1,1,\ldots,1)$, the
division is trivial --- all nodes are partitioned into one group and none in
the other.  This point defines a natural stopping criterion for the
algorithm.

The steps to detect communities in a directed network at some
intermediate stage of division therefore are:
\begin{enumerate}
\itemsep -0.5ex
\item Calculate the modularity for a subgroup.
\item If the leading eigenvector is negative or zero, the subgroup is
  indivisible.
\item If the leading eigenvector is positive, calculate the modularity of the
  entire network, assuming that the division is applied.
\item If the global modularity increases, perform the division.  If not,
  abandon the division, mark this subgroup as indivisible, and process
  another divisible subgroup.
\end{enumerate}
Repeat steps 1--4 until $Q$ reaches its maximum (or equivalently, all
subgroups are as indivisible).

To check the robustness of the results, we also apply a bottom-up algorithm
that was recently introduced by Blondel et al.~\cite{Vincent}.  Here each
network node is initially assigned to a distinct community.  Then, for each
node $i$, the gain/loss of the modularity is calculated after assigning $i$
to be in the community of one of its neighbors.  Node $i$ is then assigned to
the community that maximizes the increase of the modularity.  If there is no
increase, then node $i$ remains in its original community.  A single pass
through all network nodes defines the first stage of joinings.  This joining
step is iterated for each community in the network and continues, stage by
stage, until a maximal modularity is achieved.  This algorithm is
computationally more efficient than the previous top-down approach, but has
the disadvantage of requiring considerably more computer memory.  Both
algorithms successfully detect the same communities on simple artificial
networks, such as a collection of complete graphs that are each weakly
connected to each other.  We also compared the results of the two algorithms
when applied to the PR citation data.  We find that approximately $87.4\%$ of
the nodes are assigned to same community by both algorithms.  This result
gives a sense of the resolution with which we can define communities.

\section{Communities in the Physical Review Citation Network}
\label{sec:PR}

Our PR citation network consists of 433,452 articles published between 1893
through August 2007 with at least one citation (the nodes) and 4,370,203
total citations among these publications (the links).  To keep the scope
manageable we restrict ourselves to well-cited PR publications, defined as
those with more than 100 citations.  This restriction reduces the network to
$N=2,920$ publications and $L=11,749$ citations.  All citations to
publications outside this highly-cited set and citations from ``external'' PR
publications to this highly-cited set are excluded.  The mean degree for this
subnetwork is $2L/N\approx 8.05$.  This small value --- smaller than the mean
degree for the entire network --- stems from most of the citations to
highly-cited publications coming from articles outside this group.  We also
exclude RMP from the dataset because this journal is devoted to review
articles and the citation pattern from such articles is quite different from
other branches of PR.  Generally, RMP papers contain many more citations and
a much broader range of citations that all other PR publications.  The effect
of RMP articles is therefore to enhance citations between papers in disparate
fields.  For this reason, we exclude RMP when analyzing the community
structure of citations.

{\small \begin{longtable}
{|p{0.3in}|p{0.5in}|p{0.4in}|p{4.0in}|}
\endhead
\caption{Top-10 cited communities in the PR citation network.  Here $F$
  denotes the fraction of the highly-cited network comprised by each
  community and $N$ is the number
  publications in each community.}\label{tab-biggroups}\\
\hline rank & $F$ & $N$ & subject\\ \hline 
1 & 6.54\% & 191 & Elementary particles \\ \hline
2 & 5.99\% & 175 & Correlated electrons \\ \hline
3 & 5.86\% & 171 & Quantum information \\ \hline
4 & 5.03\% & 147 & Theories of high-Tc and type-II superconductors \\ \hline
5 & 4.97\% & 145 & Quantum diffusion, quantum Hall effect, two-dimensional melting \\ \hline
6 & 4.59\% & 134 & Statistical physics, Monte Carlo method, gauge theory, quarks \\ \hline
7 & 4.32\% & 126 & High-Tc oxides \\ \hline
8 & 4.04\% & 118 & Theories of superconductivity \\ \hline
9 & 2.95\% & 86  & Strong-coupling theory of superconductivity \\ \hline
10 & 2.60\% & 76 & Semiconductors and quantum dots \\ \hline
\end{longtable}
}

\begin{figure}[ht]
 \vspace*{0.cm}
 \includegraphics*[width=0.5\textwidth]{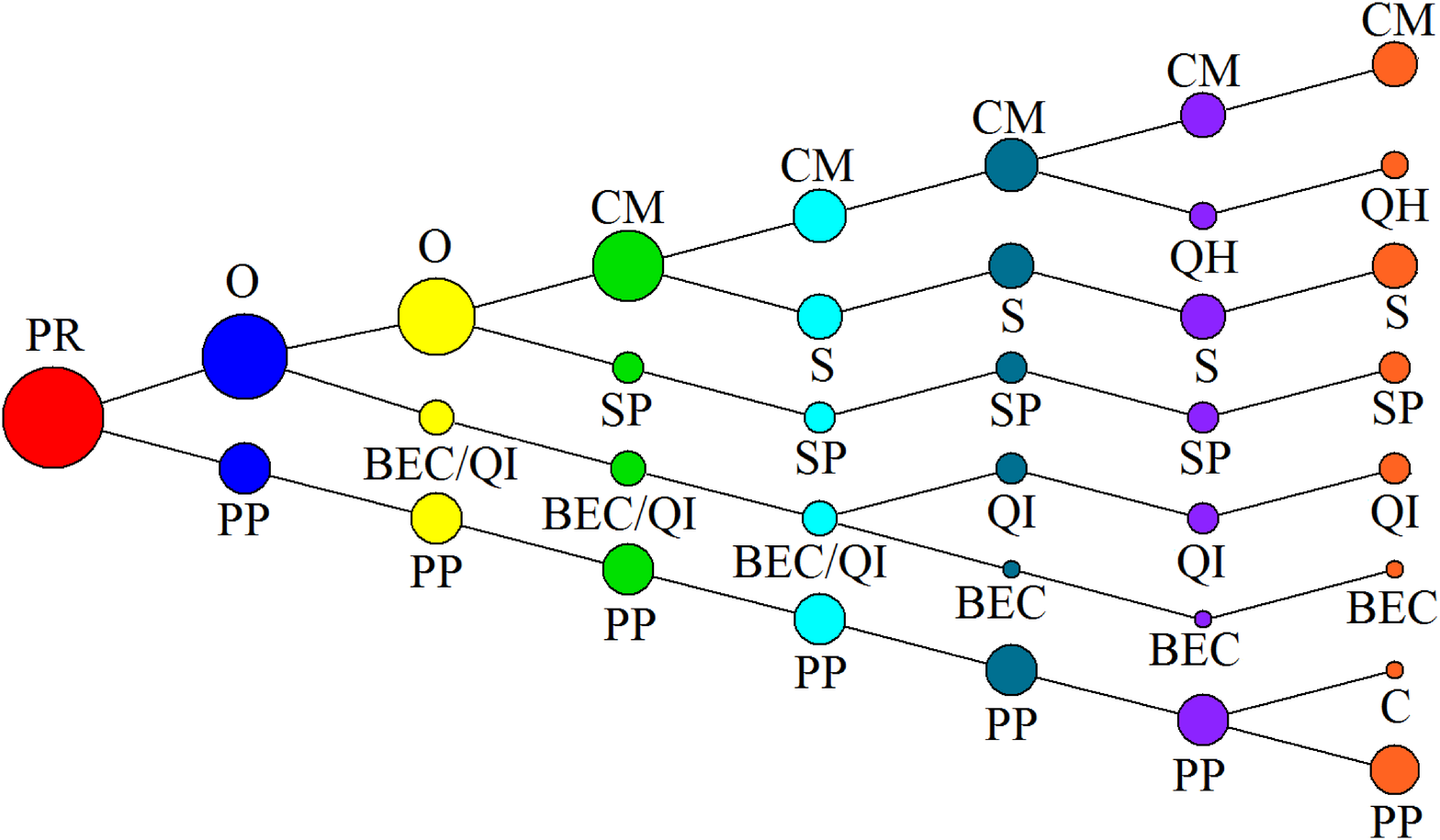}\quad\raisebox{1.2cm}{\includegraphics*[width=0.4\textwidth]{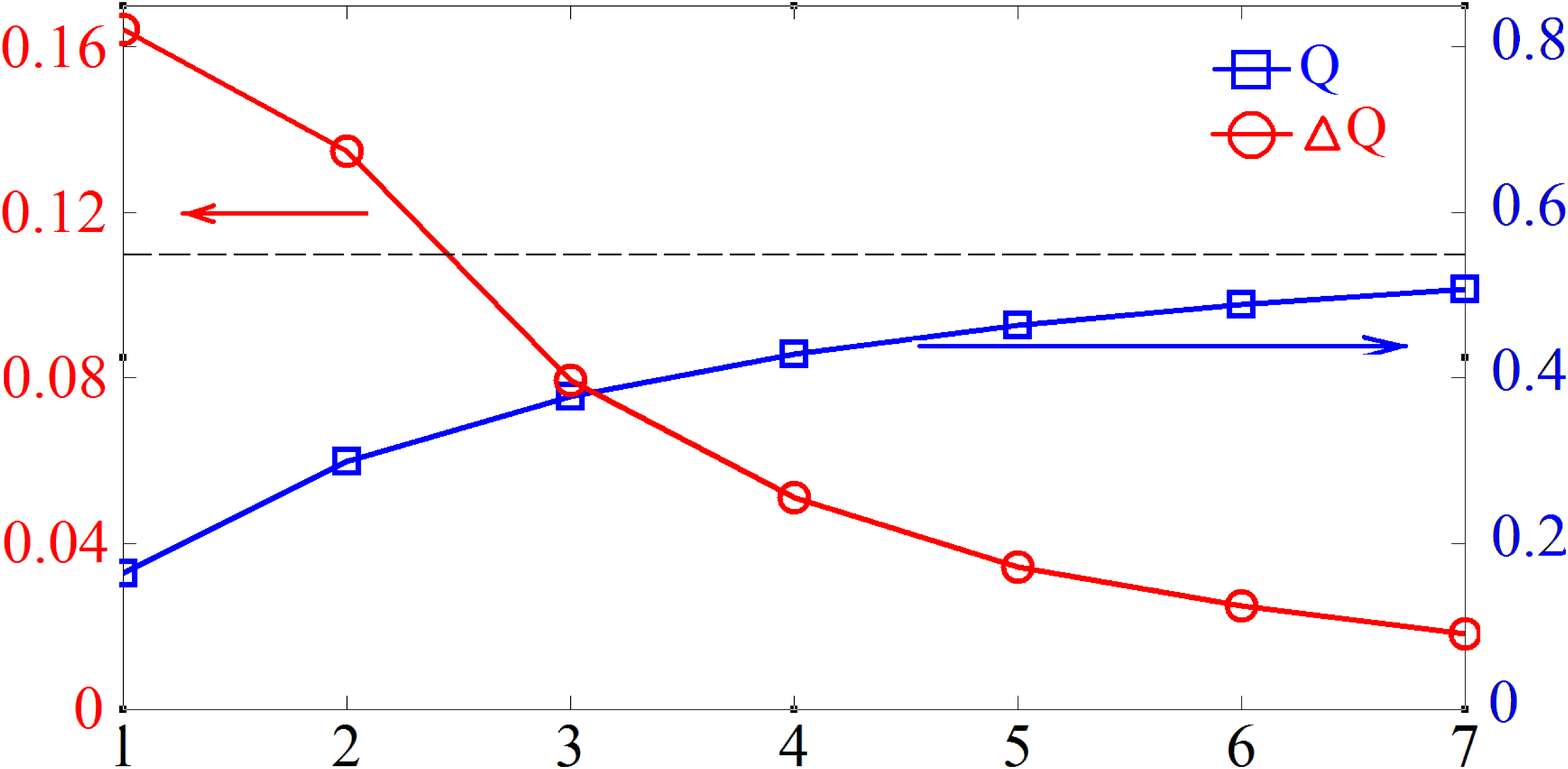}}
 \caption{(left) Evolution of the citation network after the first few
   division events.  The labels on this branching tree have the following
   correspondences: (O) other, (CM) condensed-matter physics, (QH) quantum
   Hall effect, (S) superconductivity, (SP) statistical physics (QI) quantum
   information theory, (BEC) Bose-Einstein condensation, (C) cosmology, (PP)
   particle physics. (right) The modularity $Q$ (squares) and the change of
   modularity $\Delta Q$ (circles) in the first 7 division steps, with
   different scales for $\Delta Q$ (left) and $ Q$ (right). The dotted line
   is the final modularity. } \label{steps}
\end{figure}

As modularity maximization is applied, the highly-cited (as defined above) PR
citation network divides into subgroups and the process stops when the
modularity no longer increases when additional division attempts are made.
The evolution of the initial citation network during these first few steps is
illustrated in Fig.~\ref{steps}.  After the network has been divided once, it
is, in principle, arbitrary which subnetwork to divide next.  We always
divide the subnetwork that leads to the largest increase in the overall
modularity.

\begin{figure}[!ht]
 \vspace*{0.cm}
 \includegraphics*[width=0.9\textwidth]{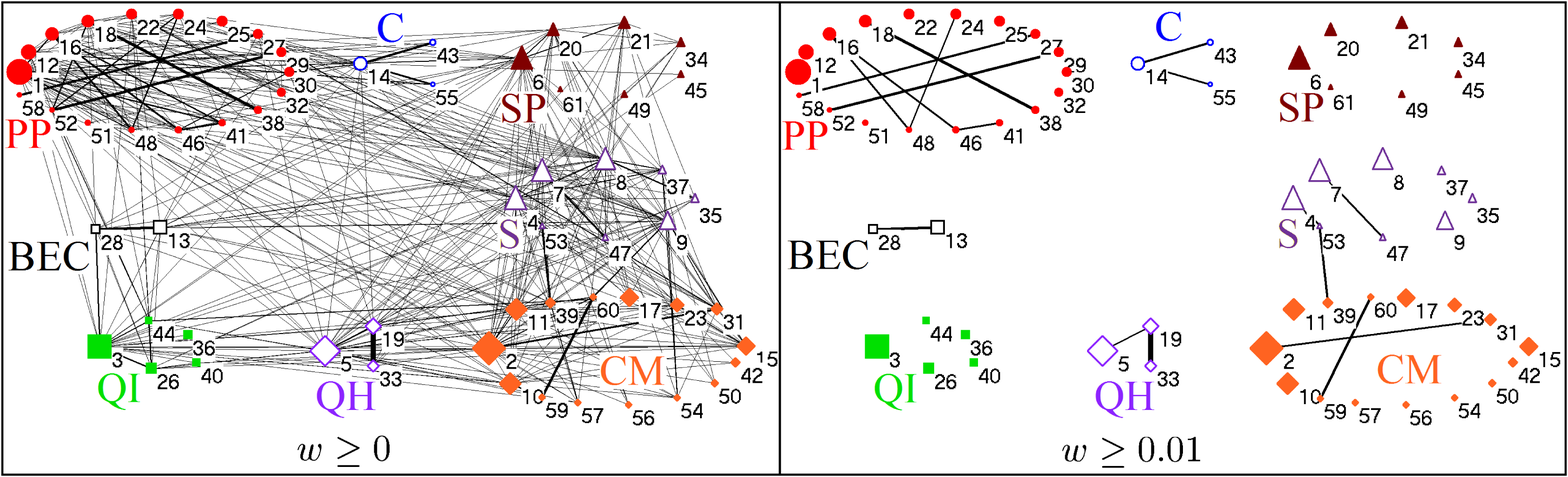}
 \caption{The 61 highly-cited citation communities with more than 5
   publications.  Symbols refer to the major groupings: particle physics
   (dots), cosmology (circles), statistical physics (solid triangles),
   superconductivity (open triangles), condensed-matter physics (solid
   diamonds), quantum Hall effect (open diamonds), quantum information
   (closed squares), and Bose-Einstein condensation (open squares).  The
   numerical labels correspond to the communities listed in
   table~\ref{tab-groups}.  The bond thickness indicates the intensity of
   citations between communities.  The left panel shows all inter-community
   links, while the right panel shows only links with weight $w \geq 0.01$
   (see text).}
\label{nv}
\end{figure}

At the end of the modularity maximization procedure, there are $274$ distinct
communities, and the network modularity is $Q=0.543$ ($Q=0.514$ if RMP papers
are included).  For these $274$ communities, the largest has $191$ members
(publications) and the smallest has only a single member.  The 10 largest
communities (listed in Table~\ref{tab-biggroups}) contain 1369 publications
and comprise $46.9\%$ of the highly-cited subnetwork.  Figure~\ref{nv} shows
the 61 communities that consist of more than $5$ publications; these are
labeled numerically corresponding to the titles of the communities that are
listed in Table~\ref{tab-groups} in the appendix.  We manually organized
these communities so that publications in the eight major groupings that
emerge after the first seven division steps of the PR network (Fig.~\ref{nv})
are spatially clustered.

To indicate the relative importance of these communities, the size of each
symbol in Fig.~\ref{nv} is proportional to the number of publications within
this community, while the width of each link is proportional to its relative
weight.  The weight $w_{ij}$ of link $ij$ is defined as $w_{ij}\equiv
{k_{ij}}/({n_in_j})$, where $k_{ij}$ is the number of citations between
communities $i$ and $j$, and $n_i$ is the number of nodes in the $i^{\rm th}$
community.  Thus $w_{ij}$ equals the fraction of all the potential
$n_i(n_i-1)/2$ links between communities that actually exist.  The average
value of the link weight for the 61 largest communities is $0.206$, while the
strongest crosslink between communities --- with weight $w_{ij}=0.056$ ---
occurs between communities 19 (fractional quantum Hall effect, theory) and 33
(fractional quantum Hall effect, experiment).  From the perspective of
modularity maximization, it is almost immaterial whether these two
communities are considered as separate or unified; joining them decreases the
modularity by only $1.36\times 10^{-5}$.

By the nature of the partitioning into communities, the links that remain
between communities at the end of modularity maximization should be weak.  In
fact, only 17 out of 393 crosslinks have a weight that exceeds 0.01 (right
side of Fig.~\ref{nv}).  Moreover, 16 out of these remaining crosslinks join
communities within the same major groupings that emerged in the initial few
division steps.  The only crosslink between different groupings joins
communities 39 (charge density waves) and 53 (Hubbard model) and it consists
of 3 citations. These connections arise from the mathematical similarity
between one-dimensional charge-density wave systems and domain wall motion in
the hall-filled Hubbard model.  Specifically, the equation of motion used in
``Incommensurate antiferromagnetism in the two-dimensional Hubbard model''
(PRL 64, 1445 (1990))~\cite{HU} and ``Continuum model for solitons in
polyacetylene'' (PRB 21, 2388 (1980))~\cite{CSP} (both from Hubbard model
community) is same as the one used for domain wall motion in ``Solitons in
polyacetylene'' \cite{SSH} (PRL 42, 1698 (1979)) and in ``Particle spectrum
in model field theories from semiclassical functional integral techniques''
\cite{PS} ({PRD 11, 3424 (1975)).  These latter two publications are both in
  charge-density wave community.

We also quantify the cohesiveness of each community by the intensity of
intra-community links.  We therefore define the intra-community link weight
$w_i \equiv {2l_i}/[{n_i(n_i-1)}]$ as the fraction of potential links that
actually exist within a community.  Here $l_i$ is the number of citations
(links) inside community $i$.  The largest value of the intra-community link
weight is $w_i=0.8$ for community 59 (atomistic metallic contacts).  The
large difference between the inter- and intra-community link weights suggest
that the partitioning that results from modularity maximization is
meaningful.

\begin{figure}[ht]
 \vspace*{0.cm}
 \includegraphics*[width=0.4\textwidth]{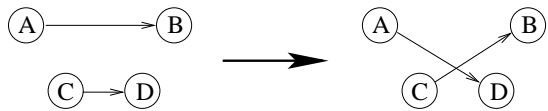}
 \caption{A single random rewiring step.  In each panel, time is increasing
   horizontally to the right.} \label{rewire}
\end{figure}

To test the significance of the communities found by modularity maximization,
we also applied this algorithm to a randomized version of the citation
network.  We construct such randomized networks by randomly rewiring links so
as to preserve the degree of each node (the number of citations to each
publication) and the time ordering of the links.  As illustrated in
Fig.~\ref{rewire}, a single rewiring step consists of first selecting two
links at random, $AB$ and $CS$, and exchanging the targets of these two links
so that the links become $AD$ and $CB$.  The rewiring step is performed only
if citing paper $A$ is earlier than cited paper $D$ and $C$ is earlier than
$B$, so as to preserve the time-ordering of the links.  This rewiring step is
applied $10 \times L$ times, where $L$ is the total number of network links.
Thus each link is rewired ten times, on average.  In this way, we preserve
the in-degree and out-degree of each network node, as well as the time
ordering of every link, but mix the global connectivity pattern.  Thus
community structure will be significantly reduced by this repeated exchange
of links.

Applying modularity maximization to these randomized networks, we find that
the modularity ranges from $0.18$ to $0.25$ for 20 different networks that
were randomized by rewiring.  These values are significantly smaller than the
modularity of $Q=0.543$ for the actual PR network.  Moreover, after the final
division of the randomized networks into communities, the number of
inter-community crosslinks is a factor 8.7 larger than that in the real PR
citation network.  Thus communities are much more cleanly defined in the PR
citation network than in the randomized networks.  This test strongly
suggests that the community structure we find in the PR citation network is a
real feature that arises from the correlations between citations.

\section{Structure of individual communities}
\label{sec:Single}

The individual communities within the PR citation network have a wide range
of structures, ranging from tightly knit to barely classifiable as a single
entity.  To illustrate this diversity, we again focus on the $61$ most
prominent PR citation communities that contain 5 or more publications
(Table~\ref{tab-groups}).  Apply modularity maximization to each community
separately, we find modularity values that range from 0.16 to 0.50 for the
communities that contain more than 25 publications.  (A modularity value of
zero does occur for 13 of the smallest communities; we ignore them because of
their small size in the following discussion.)~ As mentioned in
section~\ref{sec:MM} a modularity greater than 0.3 was empirically found
indicate community structure within a network~\cite{M_example1,M_example2}.
Among the 61 communities listed in table~\ref{tab-groups}, 23 of them have a
maximum modularity larger than 0.3.

\subsection{Most Cohesive Communities}

Let us focus on the extreme cases.  The most tightly-connected communities
are high-temperature superconductivity (high-Tc), with a modularity value of
0.194 and Bose-Einstein condensation (BEC) with a modularity value 0.217.
These two communities are illustrated in Fig.~\ref{class1}, with the titles
of the top-5 cited articles in each of them given in table~\ref{tab-group1}.
The communities are visualized by the Kamada-Kawai algorithm~\cite{KK} in
which nodes are treated as identically-charged particles and the edges are
identical springs and the algorithm arranges nodes to minimize the energy of
the system.  As is visually apparent in Fig.~\ref{class1}, these two
communities are strongly interconnected and they do not contain any visually
discernible substructure.

{\small
\begin{longtable}
{|p{0.4in}|p{3in}|p{1.0in}|p{1.2in}|}
\endhead
\caption{The top-5 cited papers in the high-Tc and BEC communities.} \label{tab-group1}\\
\hline \# cites & title of high-$T_c$ paper & authors & reference \\ \hline 

844 &Effective Hamiltonian for the superconducting $\ldots$ & Zhang and
Rice & PRB 37, 3759 (1988) \\ \hline 

643 & Superconductivity at 93K in a new mixed-phase $\ldots$ & Wu et al. &
PRL 58, 908 (1987) \\ \hline

635 & Density matrix formulation for quantum $\ldots$ & White & PRL 69, 2863
(1992) \\ \hline

634 & Effects of double exchange in magnetic crystals & de Gennes & PR 118,
141 (1960) \\ \hline

558 & Theory of high-Tc superconductivity in oxides & Emery & PRL 58, 2794
(1987) \\ \hline
\multicolumn{4}{|c|} {~} \\ 
\hline \# cites & title of BEC paper & & \\ \hline 
1119 & Bose-Einstein condensation in a gas of sodium atoms & Davis
et al. & PRL 75, 3969 (1995)\\ \hline
839 & Evidence of Bose-Einstein condensation in an $\ldots$ &
Bradley et al. & PRL 75, 1687 (1995)\\ \hline
512 & Cold bosonic atoms in optical lattices & Jaksch et al. & PRL 81, 3108 (1998)\\ \hline
388 & Vortex formation in a stirred Bose-Einstein condensate &
Madison et al. & PRL 84, 806 (2000) \\ \hline
353 & Bose-Einstein condensation of Lithium: $\ldots$ & Bradley et
al. & PRL 78, 985 (1997) \\ \hline

\end{longtable}

\begin{figure}[!ht]
      \vspace*{0.cm} \includegraphics*[width=0.95\textwidth]{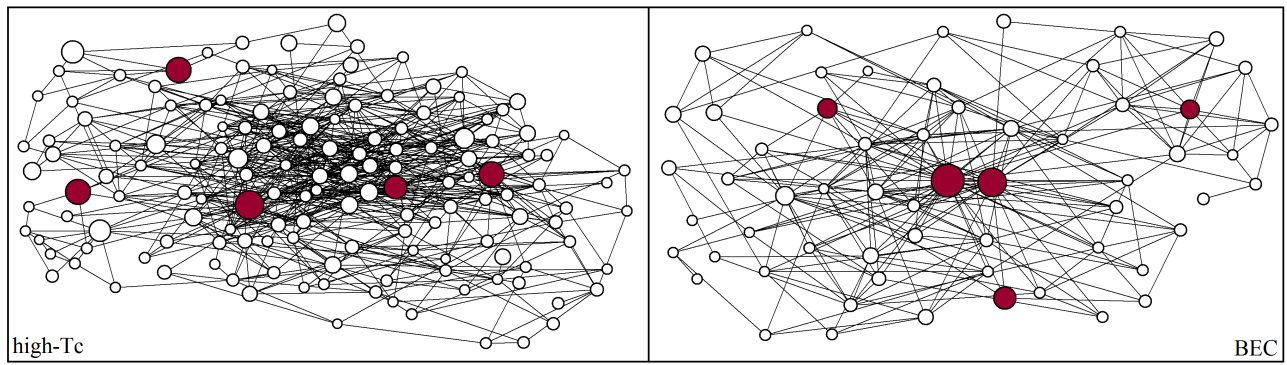}
      \caption{Visualization of communities $4$ and $13$ in Fig.~\ref{nv}:
        High-Tc and Bose-Einstein condensation, respectively.  The top-$5$
        cited papers in each community are denoted by dots.  The size of each
        symbol is proportional to its number of citations.} \label{class1}
\end{figure}

\subsection{Least Cohesive Community}

On the other hand, the communities with the highest modularity values are
quite different in character.  If these communities were treated as isolated
entities in the absence of the rest of the network, modularity maximization
would divide them into still smaller communities.  The resulting large value
of the modularity that is found after these communities are divided suggests
that the substructure within each community is significant.  The
least-cohesive such example is community 6 (statistical physics, Monte Carlo
methods, gauge theory, and quarks), with modularity 0.498.  This community is
shown in Fig.~~\ref{theory} and table~\ref{tab-group3} lists the top-$5$
cited papers for this community.

It may seem surprising, at first sight, that papers such as ``Thermal
fluctuations, quenched disorder, phase transitions, and transport in type-II
superconductors'' and ``Dynamic scaling of growing interfaces'' are in the
same community as ``Confinement of quarks''.  The origin of the connection
between these publications is illustrated in Fig.~\ref{theory}, where the
community appears to have three distinct modules.  Here we manually arrange
the nodes so that the connections between these modules are highlighted.
Statistical physics papers (SP) appear in the large ellipse are on the left,
Monte Carlo and gauge theory papers (MC) are in the middle, and quark-related
papers (Q) are on the right.  There are 22 links between the SP and MC
modules; for visual clarity, only the 6 most significant links (in which one
or both of the link ends attach to a node with more than 500 citations) are
shown. Also shown are all of the 6 links between the MC and Q modules.

{\small
\begin{longtable}
{|p{0.4in}|p{3in}|p{1.2in}|p{1.3in}|}
\endhead
\caption{Statistical physics, Monte-Carlo method, gauge theory, and quarks}\label{tab-group3}\\
\hline \# cites & title & authors & reference \\ \hline
1009 & Dynamic scaling of growing interfaces & Kardar et al. & PRL 56, 889 (1986) \\ \hline
852 & Absence of ferromagnetism or antiferromagnetism $\ldots$ & Mermin and
Wagner & PRL 17, 1133 (1966) \\ \hline
783 & Thermal fluctuations, quenched disorder, phase $\ldots$ & Fisher et al.
& PRB 43, 130 (1991) \\ \hline
660 & Crystal statistics.I. a two-dimensional model with $\ldots$ & Onsager & PR 65, 117 (1944) \\ \hline
655 & Confinement of quarks & Wilson & PRD 10, 2445 (1974) \\ \hline
\end{longtable}
}

\begin{figure}[!ht]
 \vspace*{0.cm}
 \includegraphics*[width=0.75\textwidth]{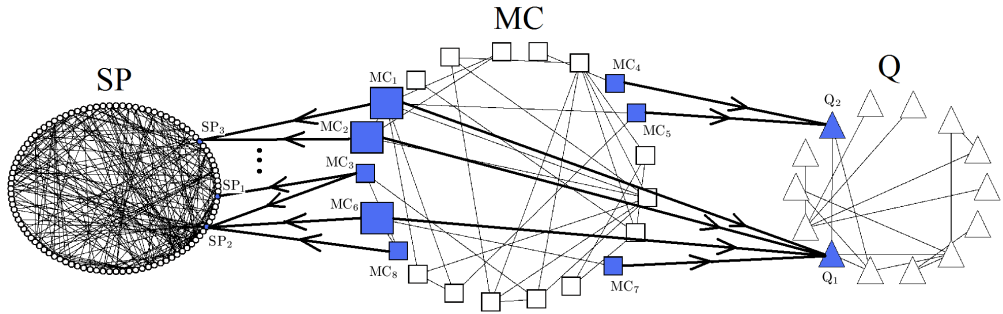}
 \caption{Community 6 in Fig.~\ref{nv}: statistical physics, Monte Carlo
   methods, gauge theory, and quarks, that is divided into the modules of
   statistical physics (SP), Monte Carlo methods and gauge theory (MC), and
   quarks (Q).  Filled symbols denote the labeled nodes in
   table~\ref{tab-crossgroup1}.  The larger nodes directly join the three
   modules through their citations.}
\label{theory}
\end{figure}

To highlight the connections between these modules, we focus on the subset of
citations, marked by thick lines in Fig.~\ref{theory}, that join publications
across all three modules.  We label the nodes at the ends of these links by
SP$_1$--SP$_3$, MC$_1$--MC$_8$, Q$_1$ and Q$_2$, respectively, for
statistical physics, Monte Carlo, and quark publications.  The subscripts
order these nodes by their number of citations.  Among these highly-cited
nodes, MC$_1$, MC$_2$, and MC$_6$ are exceptional because they directly link
modules SP and Q.  In MC$_1$: ``New Monte Carlo technique for studying phase
transitions'' (PRL 61, 2635 (1988)) and MC$_2$: ``Optimized Monte Carlo data
analysis'' (PRL 63, 1195 (1989)), Ferrenberg and Swendsen introduced an
efficient Monte Carlo technique to extract information at many temperatures
from data that is generated at a single temperature.  This technique is
especially useful in the Ising model and in lattice gauge theory, where
significant computational resources are needed.  The former is a model of
ferromagnetism and is widely used in statistical physics.  The latter is an
important tool for studying the confinement of color-charged particles (such
as quarks).  Publication MC$_6$ ``Order and disorder in gauge systems and
magnets'' (PRD 17, 2637 (1978)) showed that the renormalization-group
equations of four-dimensional gauge theories (used in Q$_1$) have the same
structure as those of two-dimensional spin systems (discussed in SP$_2$).
Because of the shared techniques (large-scale Monte Carlo simulations) and
similar mathematical structure (the analogies between gauge theories and spin
models), the SP and Q modules, which might seem to belong to disparate
physics fields, are actually connected by module MC.

{\small
\begin{longtable}
{|p{0.3in}|p{0.4in}|p{3in}|p{1.5in}|p{1.3in}|}
\endhead
\caption{Bridge nodes for statistical physics, Monte Carlo method, gauge theory, and quarks community}\label{tab-crossgroup1}\\
\hline label & \# cites & title & authors & reference \\ \hline
SP$_1$ & 852 & Absence of ferromagnetism or antiferromagnetism $\ldots$
& Mermin and Wagner & PRL 17, 1133 (1966) \\ \hline
SP$_2$ & 602 & Renormalization, vortices, and $\ldots$ & Jose and Kadanoff & PRB 16, 1217 (1977)\\
\hline
SP$_3$ & 172 &  Bounded and inhomogeneous Ising models: $\ldots$ &
Ferdinand and Fisher & PR 185, 832 (1969) \\ \hline
MC$_1$ & 404 &  New Monte Carlo technique for studying phase $\ldots$ &
Ferrenberg and Swendsen & PRL 61, 2635 (1988) \\ \hline
MC$_2$ & 301 &  Optimized Monte Carlo data analysis &  Ferrenberg and
Swendsen & PRL 63, 1195 (1989)\\ \hline
MC$_3$ & 135 &  Monte Carlo study of the planar spin model & Tobochnik
and Chester & PRB 20, 3761 (1979) \\ \hline
MC$_4$ & 129 &  High-temperature Yang-Mills theories and $\ldots$ &
Appelquist and Pisarski & PRD 23, 2305 (1981) \\ \hline
MC$_5$ & 125 &  Critical properties from Monte Carlo coarse $\ldots$ &
Binder & PRL 47, 693 (1981) \\ \hline
MC$_6$ & 120 &  Order and disorder in gauge systems and magnets &
Fradkin and Susskind & PRD 17, 2637 (1978)\\ \hline
MC$_7$ & 104 &  Impossibility of spontaneously breaking local $\ldots$
& Elitzur & PRD 12, 3978 (1975)\\ \hline
MC$_8$ & 103 &  Topological excitations and Monte Carlo $\ldots$ &
DeGrand and Toussaint & PRD 22, 2478 (1980)\\ \hline
Q$_1$ & 655 &  Confinement of quarks & Cardona et al.& PR 154, 696 (1967) \\ \hline
Q$_2$& 108 &  Lattice models of quark confinement at high $\ldots$ &
Susskind & PRD 20, 2610 (1979) \\ \hline
\end{longtable}
}

The example of the statistical physics, Monte-Carlo method, gauge theory, and
quarks community suggests that the reducibility of a community by modularity
maximization depends on the nature of its embedding in the rest of the
network.  When this particular community is isolated it is reducible, but it
is irreducible when considered part of the entire citation network.  It is
worthwhile to understand this feature in an idealized example.  Thus consider
two complete graphs of 100 nodes each in which the fraction of all possible
links between these two graphs is $p$.  When this twinned complete graph is
isolated, a value of $p<0.9$ is sufficient to induce modularity maximization
to divide this network into its two constituent complete graphs.  However, if
the double-graph system is embedded in a larger network, the value of $p$
needed for division to occur quickly decreases with the size of the larger
network.

\begin{figure}[ht]
 \vspace*{0.cm}
\raisebox{1.6cm}{ \includegraphics*[width=0.3\textwidth]{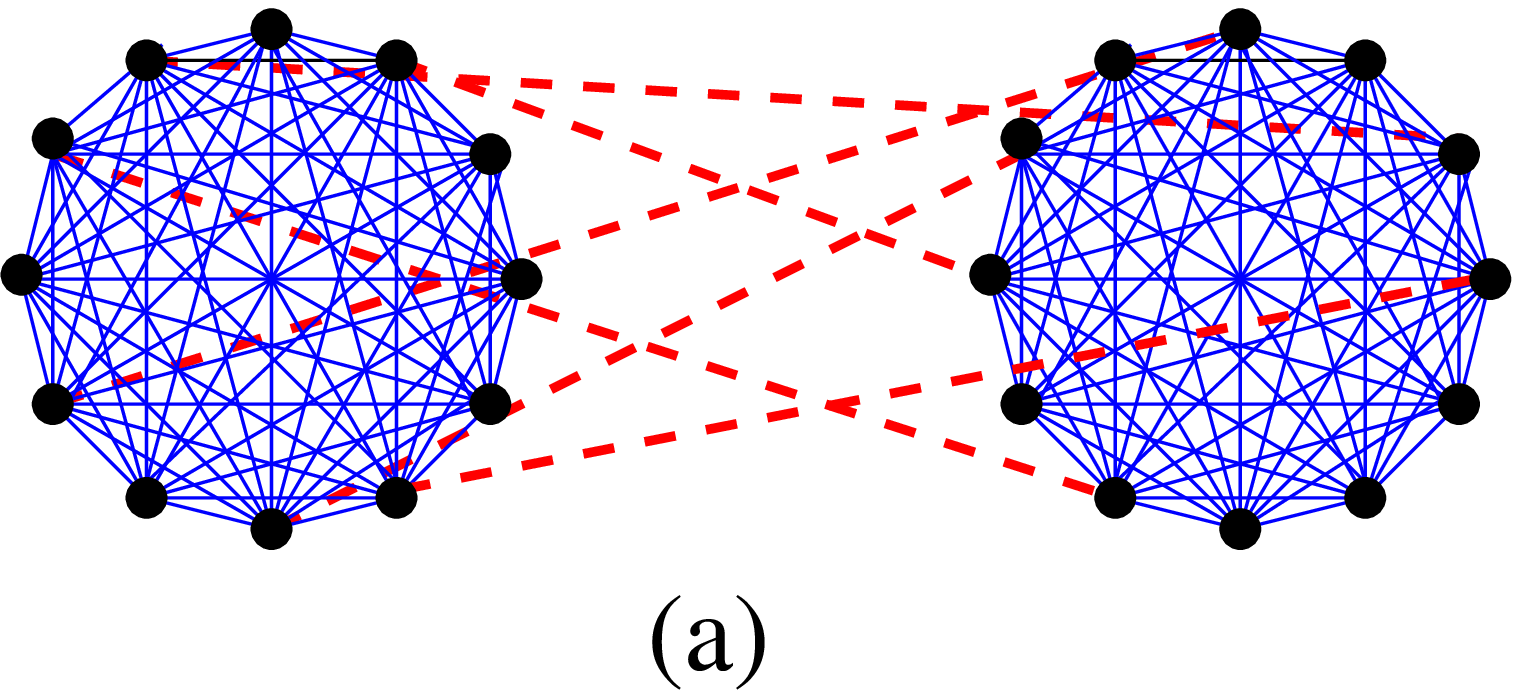}}\hskip 1.5in\includegraphics*[width=0.3\textwidth]{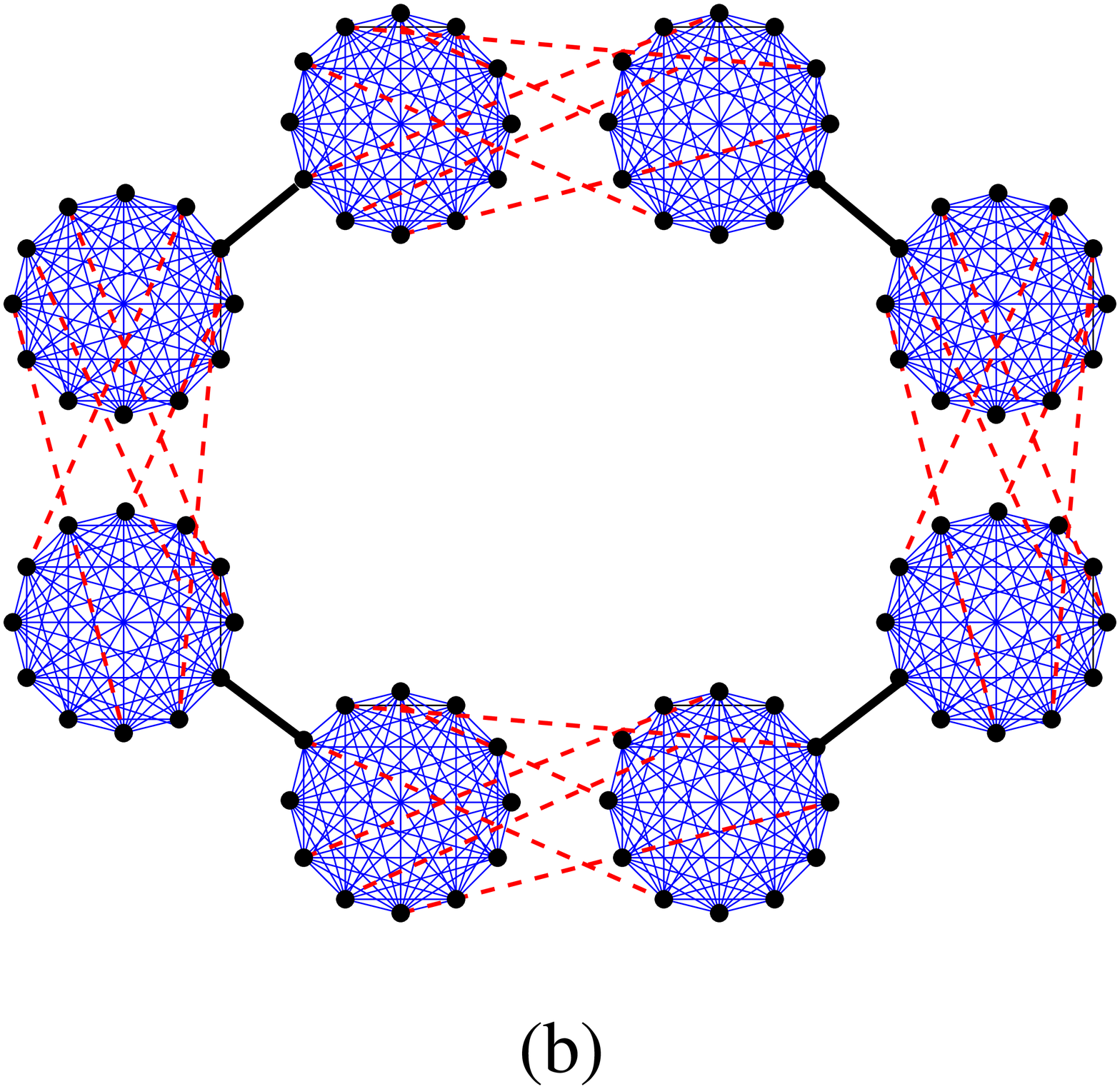}
\caption{(a) A twinned complete graph that consists of two complete graphs,
  with a fraction $p$ of the possible links (dashed) between them also
  present.  (b) Four twinned complete graphs interconnected by a single link
  (heavy solid lines). }
\label{double}
\end{figure}

For example, consider four identical copies of the twinned complete graph, in
which each such graph is connected to another by a single link
(Fig.~\ref{double}).  In this case, modularity maximization for the entire
network immediately breaks all the single links, as one would naturally
expect.  However, the individual complete graph twins remain joined unless the
value of $p$ is less than 0.2.  Thus there is a wide range of $p$ values for
which an isolated twinned complete graph will be split by modularity
maximization, but will remain intact if this same twinned complete graph is
part of a larger network.

\section{Time Evolution}
\label{sec:TV}

The community structure presented in the previous sections represents an
integrated view of the PR citation network over the 114-year period
1893--2006.  Physics is an evolving discipline, however, and it is revealing
to study how the community structure of the citation network evolves over
time.  To this end we partitioned all PR publications into eight decadal sets
by year of publication: 1927--1936, 1937--1946, $\ldots$, 1997--2006.  Our
choice of decades was dictated by 2006 representing the last year for which
complete citation data was available.  We also restrict our analysis to
roughly the top-3000 cited papers in each decade.  More precisely, we adjust
the threshold for number of citations to make the total papers in each decade
to be as close to 3000 as possible (the actual number in each decade ranged
from 2751 to 3046).  Because the Physical Review has been growing roughly
exponentially with time since 1893 (with somewhat slower growth after World
War II than before~\cite{SID110}), each decadal snapshot represents a
different fraction of the total number of publications in this period.

Table~\ref{tab-decade} lists basic information about the publications in each
of the eight decades.  The fraction of publications that we analyze (column
$P$) becomes quite small in the last four decades so that our results from
later decades are biased toward highly-cited papers.  We did not analyze the
period before 1927 because the number of papers is too small to resolve them
into meaningful communities.  We categorize these communities as belonging to
PRA/E (atomic, molecular, optical, statistical), PRB (condensed-matter), PRC
(nuclear), or PRD (particle).  We treat Physical Review E (statistical
physics) and Physical Review A (atomic, molecular, optical physics) together,
since PRE was split off from the latter only in 1990.  To assign communities
to categories, we adopt the following procedure: after 1970, papers that
appeared in PRA/E, PRB, PRC, and PRD can be unambiguously be assigned their
subject category.  For papers in the letters section (PRL) or for papers
published before 1970, we count the fraction of their citations that come
from PRA/E, PRB, PRC, and PRD.  We then assign a paper the category of the
plurality of its citations.  Finally, a community can be categorized by the
plurality of all the citations to its constituent publications.

There are some ambiguities with this procedure that should be noted.  First,
we can only use citing papers after 1970 (that appear in the four sections of
Physical Review) to categorize papers that were published before 1970.  Thus
citations from papers that were published before 1970 are not used.  Second,
the categorization of some communities is not definitive.  For example, the
category fractions for community 401 (Pion \& nucleon interactions) is
$24\%$, $11\%$, $34 \%$, and $32\%$, respectively, for PRA/E, PRB, PRC, and
PRD.  Since PRC has the largest share, the community is categorized as being
part of PRC.  Fortunately, for the 109 decadal communities in
Fig.~\ref{decades}, 101 of them have more than a 50\% share in one category.

{\small
\begin{longtable}
{|p{0.6in}|p{0.2in}|p{0.4in}|p{0.25in}|p{0.25in}|p{0.25in}|p{0.45in}|p{0.35in}|p{0.35in}|p{0.35in}|}
\endhead
\caption{The eight decadal publication sets 1927--36 to 1997--2006.  Here $T$
  is the citation threshold for inclusion in the dataset; $N$ is number of PR
  papers in each decade, $M$ is number of decadal papers analyzed,
  $P={M}/{N}$, and $Q$ is the maximum modularity in each decade.  The last 4
  columns are the fraction of papers that belong to the categories PRA/E,
  PRB, PRC, and PRD (see text).}
  \label{tab-decade}\\
\hline
Decades & T & $N$ & $M$ & $P$ & $Q$ &PRA/E &PRB &PRC &PRD\\
\hline
1927-1936 & 2 & 3908 & 2751 & $70\%$ & 0.50 & $84.1\%$ & $15.9\%$ &
$0\%$ & $0\%$ \\ \hline
1937-1946 & 1 & 3530 & 3007 & $85\%$ & 0.51 & $39.9\%$ & $0\%$ &
$51.4\%$ & $8.7\%$ \\ \hline
1947-1956 & 14 & 12692 & 2994  & $24\%$ & 0.40 & $9.6\%$ & $27.3\%$
& $43.0\%$ & $20.1\%$ \\ \hline
1957-1966 & 25 & 20642 & 3046 & $15\%$ & 0.52 & $10.3\%$ & $48.7\%$
& $11.9\%$ & $29.2\%$ \\ \hline
1967-1976 & 34 & 43628 & 2982 & $7\%$ & 0.56 & $12.1\%$ & $57.1\%$ &
$0\%$ & $30.8\%$ \\ \hline
1977-1986 & 44 & 54475 & 2950 & $5\%$ & 0.59 & $6.7\%$ & $68.2\%$ &
$0\%$ & $25.1\%$ \\ \hline
1987-1996 & 59 & 103774 & 2997 & $3\%$ & 0.61 & $15.7\%$ & $79.6\%$
& $0\%$ & $4.6\%$ \\ \hline
1997-2006 & 42 & 151693 & 2954 & $2\%$ & 0.64 & $41.5\%$ & $31.6\%$
& $3.7\%$ & $23.2\%$ \\ \hline
\end{longtable} }

Figure~\ref{decades} illustrates these time-resolved communities decade by
decade since 1927, showing both connections within each decade and
connections between communities in different decades.  Along the vertical
axis, the communities are arranged according to the categories PRA/E, PRB,
PRC, and PRD.  There are 109 decadal communities with more than 50
publications and their subjects are listed in tables \ref{tab-decade-1} to
\ref{tab-decade-4} (see appendix).  The node size again is proportional to
the number of publications in each community.  For visual clarity, we do not
show links with weights $w_{ij} \leq 0.001$.  The strongest connections occur
between communities within the same decade and between communities in
consecutive decades.  These short-range temporal connections are natural to
expect because of the average citation lifetime is only 6 years~\cite{SID110}
and because intellectually-close publications may be in neighboring decadal
datasets.  Because of the arbitrariness of the partitioning into decades, a
single subfield may appear as multiple communities in adjacent decades.
Consequently, links between communities in adjacent decades should be viewed
as equivalent to links within a given decade.  An indication that two
communities in adjacent decades are really a single community is that their
topics are similar and there is an above-average number of interconnecting
links between them.  Thus, for example, the communities ``Conductance
fluctuations/scaling'' (703 in Fig.~\ref{decades}) from 1977-1986 and
community ``Conductance fluctuations'' (labeled 814) from 1987-1996 are
likely part of the same community.  The intensity of citations (the number
divided by the product of the two community sizes) between them is 10.2 times
larger than the average intensity of citations between any two communities.

\begin{figure}[H]
 \vspace*{0.cm}
 \includegraphics*[width=0.9\textwidth]{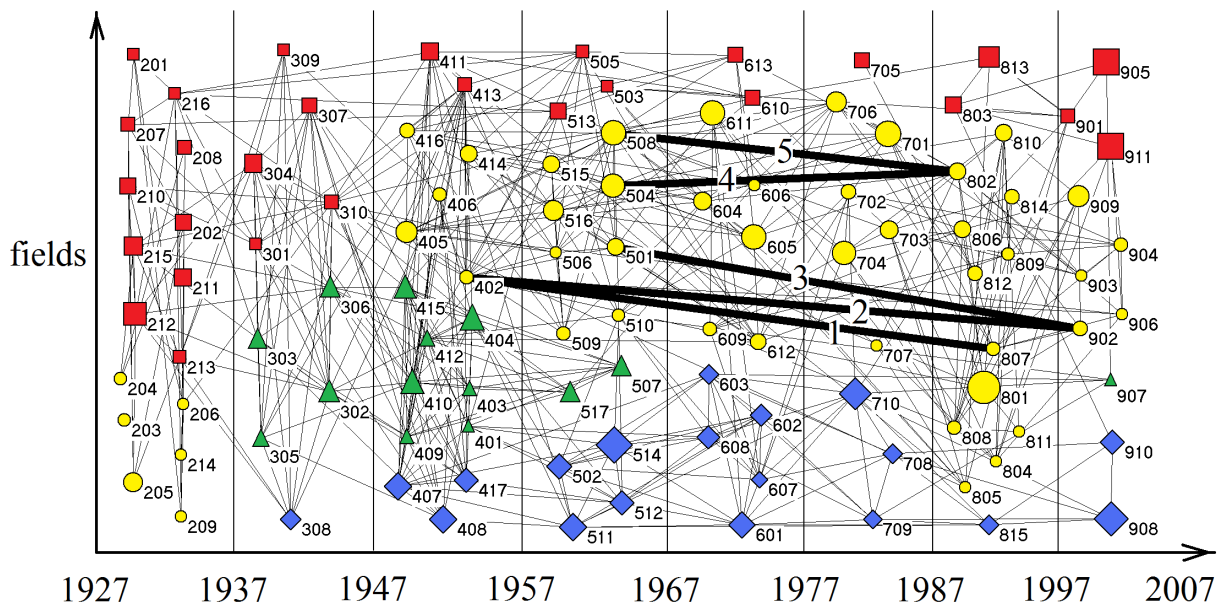}
 \caption{Citation communities decade by decade.  Communities are arranged
   from top to bottom categories by PRA/E; (squares), PRB (circles), PRC
   (triangles), or PRD (diamonds).  The thick lines indicate the top-5
   long-range temporal connections between communities. } \label{decades}
\end{figure}

There are a variety of interesting anomalies in the decadal data.  During the
decades 1927--36 and 1997--2006 atomic physics (PRA) was the largest field.
From 1927--36, the activity in this field was likely due to the revolution in
quantum mechanics.  Between 1997--2006, the upsurge in atomic physics is due
to the developments in Bose-Einstein condensation and in quantum information
theory.  Condensed-matter physics (PRB) is relatively stable throughout all
eight decades, but with two noteworthy features.  First, in 1987--96,
condensed-matter physics represented $79.6\%$ of all highly-cited papers in
this decade (defined as more than 59 citations, see table~\ref{tab-decade}).
Moreover $40.1\%$ of all papers were related to high-temperature
superconductivity.  Second, during 1937--46, none of the communities shown in
Fig.~\ref{decades} belong to condensed-matter physics, while $51.4\%$ of all
papers in this decade were in nuclear physics.

The fraction of highly-cited nuclear physics publications reached a maximum
during 1937--46, remained significant for the following decade, but then
almost completely disappeared from the most highly-cited papers after 1967
(only 1 community in 1997--2006 can be categorized as belonging to PRC).
Last, particle physics (PRD) has had a fraction of between $20\%--30\%$ of
the highest-cited papers, except for 1987--1996 ($4.6\%$).  The small share
for this field between 1987--96 was caused by the upsurge in high-temperature
superconductivity, Bose-Einstein condensation, and quantum information.
Amazingly, these three topics comprised $63.1\%$ of all highly-cited papers
in this decade.

Another interesting feature that is visible in Fig.~\ref{decades} is the
existence of a very small number of long-range links in time.  In fact, there
are only five links that span more than two decades in time and have a link
weight greater then 0.004 (table~\ref{tab-longlink}).  These long-range links
are indicative of significant scientific developments.  For example, links 1,
2, and 3 in Fig.~\ref{decades} can be traced to the phenomenon of colossal
magnetoresistance (CMR).  Magnetoresistance is the change in electrical
resistance of a material when a magnetic field is applied.  In 1951, Zener
explained the magnetoresistance of manganese oxides by the double exchange
mechanism~\cite{zener}.  It was only in the mid-1980s, that advances in
nanotechnology allowed experimentalists to create the types materials
proposed in earlier theoretical studies.  In fact, these materials exhibited
``giant'' magnetoresistance (GMR), a phenomenon that led to the development
of modern hard-disk industry and to the Nobel prize in physics to Fert and
Grunberg in 1988 for their discoveries in this field.  Subsequent
high-precision fabrication techniques led to the even more extreme phenomenon
of ``colossal'' magnetoresistance (CMR).  The three long-range links
mentioned above owe their existence to the delayed experimental renaissance
of theoretical ideas.

Links 4 and 5 can be attributed to two famous condensed-matter physics papers
--- ``Self-consistent equations including exchange and correlation Effects''
by Kohn and Sham (KS) and ``Inhomogeneous electron gas'' by Hohenberg and
Kohn (HK).  Link 4 consists of 33 citations between the communities
``Correlated electron systems'' and ``Pseudopotentials'', and 26 out of them
are solely based on KS.  This publication is, in fact, the most-cited in all
Physical Review, with 4849 {\em internal\/} citations (citations from other
PR publications to KS) as of the fall of 2009.  Similarly link 5 consists of
25 citations between communities ``Many-body systems'' and
``Pseudopotentials'', of which 21 of them are due to HK.  This publication is
the second most-cited PR paper.  These two papers continue to be heavily
cited more than 40 years after publication because of the wide usage of the
approximation methods for dealing inhomogeneous and interacting electron
systems.

{\small
\begin{longtable}
{|p{0.5in}|p{2.5in}|p{2.5in}|}
\endhead
\caption{The top-5 long-range links. }
\label{tab-longlink}\\ \hline
link & community name  & community  name\\\hline
1 &NMR (1947-1956) & Giant magnetoresistance (1987-1996) \\ \hline
2 &NMR (1947-1956) & Manganite (1997-2006)\\ \hline
3 &Quantum magnetism (1957-1966) &``Manganite'' (1997-2006) \\ \hline
4 &Correlated electron systems (1957-1966) & Pseudopotentials (1987-1996) \\ \hline
5 &Many-body systems (1957-1966) & Pseudopotentials (1987-1996)  \\ \hline
\end{longtable}
}

\section{Summary}
\label{sec:Summary}

The recent availability of large citation datasets has made it feasible to
study properties of citation networks that were inaccessible even one decade
ago.  One of the goals of this work is to understand the structure of the
citation network of the most prominent US archival physics journal --- the
Physical Review (PR) family of journals.  Our goal was to determine basic
properties of the subfields of physics --- their importance, their
interconnections, and their evolution --- by focusing on citations rather
than on the content of the publications themselves.  To identify subfields
within the PR citation network, we used an algorithm that repeatedly attempts
to partition the network into well-defined communities by maximizing a
measure known as the modularity.  Using this approach, the PR citation
network was found to possess an underlying community structure that
corresponds to clearly identifiable subfields.

We also we studied the structure of the individual communities that were
identified by modularity maximization.  By treating each community as an
independent network and again applying modularity maximization, we found that
these individual communities were diverse in their structures, ranging from
tightly focused to barely being identifiable as a community.  The most weakly
defined communities consist of a small number of well-defined modules that
are typically linked by publications that emphasize techniques (as in the
example of the Statistical physics, Monte Carlo method, gauge theory, and
quarks community discussed in Sec.~\ref{sec:Single}.  These weakly defined
communities --- when isolated from the rest of the network --- will be
divided further by modularity maximization.  However, modularity maximization
leaves these communities intact when they are considered as part of the
entire network, so that some of the communities do not appear coherently
organized around a single theme.

We also studied the time evolution of the citation network and found five
exceptionally long-lived links between subfields that are separated by more
than two decades in time (compared to the average PR citation age of
approximately 6 years).  These long-range links arise from one of two
mechanisms: either (i) a long delay between theoretical insights and the
development of experimental methods to implement these ideas (links 1, 2, and
3 in Fig.\ref{decades}), or (ii) the introduction of a widely-used new method
(links 4 and 5).  We also uncovered a number of anomalies in the evolution of
the four major categories of Physical Review (PRA/E, PRB, PRC, and PRD)
throughout the decades 1927--36 until 1997--2006 that can be traced to major
historical events.  The more prominent such examples include: (i) the
flowering of nuclear physics in the period just before and just after WWII,
where the share of PRC (nuclear physics) publications was maximal, (ii) the
discovery of high-temperature superconductivity, where the share of
highly-cited PRB (condensed-matter physics) publication rose dramatically in
the 1980s, and (iii) the burgeoning of quantum information theory over the
past decade, leading to a sharp increase in publications in PRA (atomic,
molecular, optical physics).

\acknowledgments{We are grateful to Mark Doyle and Paul Dlug from the APS for
  providing the Physical Review citation data.  We are also grateful for
  financial support from NSF grants DMR0535503 and DMR0906504 .}

\newpage

\section*{Appendix}

{\small
\begin{longtable}
{|p{0.15in}|p{0.35in}|p{0.2in}|p{0.3in}|p{2.0in}|p{0.2in}|p{0.15in}|p{0.35in}|p{0.2in}|p{0.3in}|p{2.0in}|}
\endhead
\caption{Top 61 communities in the PR citation network.  Here $R$ denotes the
  citation rank, $F$ and $N$ are fraction and the number of all highly-cited
  publications in each community, and $Q$ is the local modularity for each
  community.  Communities with modularity values greater than 0.4 are
  generally multi-themed, as described in the text, while modularity values
  of zero occur only for the smallest communities. }\label{tab-groups}\\
\hline   $R$ & $F$ & $N$ & $Q$
&\phantom{Quantum diffusion, quantum Hall}&~~ &  R & F & N
&\phantom{Quantum diffusion, quantum Hall}&\\ \hline

1 & 6.54\% & 191 & .371& Elementary particles&& 32 & 0.75\% & 22 & .312& Cross
sections \\ \hline

2 & 5.99\% & 175 & .240& Correlated electrons&& 33 & 0.72\% & 21 &
0& \hbox{Fractional quantum Hall effect:} experiment\\ \hline

3 & 5.86\% & 171 & .398 &Quantum information && 34 & 0.68\% & 20 & .318& Self
organized criticality \\ \hline

4 & 5.03\% & 147 & .194& \hbox{Theories of high-Tc and type-II} superconductors
&& 35 & 0.68\% & 20 & .279& Magnetism \\ \hline

5 & 4.97\% & 145 & .436& \hbox{Quantum diffusion, quantum Hall} effect and
two-dimensional melting && 36 & 0.65\% & 19 & .223& High-harmonic
generation \\ \hline

6 & 4.59\% & 134 & .498& \hbox{Statistical physics, Monte Carlo} method, gauge
theory, and quarks && 37 & 0.65\% & 19 & 0& Fermi surface \\ \hline

7 & 4.32\% & 126 & .424& High-Tc oxides && 38 & 0.62\% & 18 & 0& Quarks and
infinite momentum \\ \hline

8 & 4.04\% & 118 & .386& Theories of superconductivity && 39 & 0.55\% & 16
& .234& Charge density waves \\ \hline

9 & 2.95\% & 86  & .441& \hbox{Strong-coupling theory of} superconductivity && 40
& 0.55\% & 16 & .191& Self-induced transparency \\ \hline

10 & 2.60\% & 76 & .487& Semiconductors and quantum dots && 41 & 0.48\% &
14 & .308& Weak interactions \\ \hline

11 & 2.50\% & 73 & .426& Material Science && 42 & 0.45\% & 13 & .165& Scanning
tunneling microscope \\ \hline

12 & 2.33\% & 68 & .366& Quantum field theory && 43 & 0.45\% & 13 &
.168& Astrophysics \\ \hline

13 & 2.05\% & 60 & .217& B.E. condensation && 44 & 0.45\% & 13 & 0& Quantum
optics \\ \hline

14 & 1.99\% & 58 & .251& Cosmology && 45 & 0.41\% & 12 & .223& Stochastic
resonance \\ \hline

15 & 1.82\% & 53 & .481& Metals and alloys && 46 & 0.38\% & 11 & .286& Top quark
\\ \hline

16 & 1.75\% & 51 & .314& \hbox{Symmetry breaking and gauge} theories && 47 &
0.34\% & 10 & .238& Pinning in superconductors \\ \hline

17 & 1.71\% & 50 & .320& Giant magnetoresistance && 48 & 0.34\% & 10 &
.078& $e^{+}e^{-}$ annihilation \\ \hline

18 & 1.51\% & 44 & .322& High energy collisions && 49 & 0.34\% & 10 &
.156& Synchronized chaos \\ \hline

19 & 1.40\% & 41 & .206& \hbox{Fractional quantum Hall effect:} theory&&  50
& 0.27\% & 8  & 0& Surface transitions \\ \hline

20 & 1.34\% & 39 & .381& Equilibrium statistical mechanics &&51 & 0.27\% &
8 & 0& Atomic collision interactions \\ \hline

21 & 1.23\% & 36 & .161& Diffusion limited aggregation &&  52 & 0.24\% & 7
& 0& Elastic scattering \\ \hline

22 & 1.06\% & 31 & .447& Quantum mechanics && 53 & 0.24\% & 7  & 0& Hubbard
model \\ \hline

23 & 1.03\% & 30 & .393& Crystal structure && 54 & 0.24\% & 7  & .223& Optical
properties of metals \\ \hline

24 & 0.99\% & 29 & .361& Nucleon-nucleon interactions && 55 & 0.24\% & 7
& .198& Reheating after inflation \\ \hline

25 & 0.99\% & 29 & .305& Neutrino oscillations && 56 & 0.24\% & 7  &
0& Magnetic anisotropy \\ \hline

26 & 0.99\% & 29 & .326& Quantum teleportation && 57 & 0.24\% & 7  &
0& Transport in disordered systems \\ \hline

27 & 0.92\% & 27 & .428& \hbox{Collective electronic properties/~~~~~} nuclear fission
&& 58 & 0.24\% & 7  & .210& Nuclei \\ \hline

28 & 0.89\% & 26 & .278& Quantum entanglement && 59 & 0.21\% & 6  &
0& Atomistic-sized metallic contacts \\ \hline

29 & 0.82\% & 24 & .282& Nuclear collision && 60 & 0.21\% & 6  & 0& Carbon
nanotubes \\ \hline

30 & 0.79\% & 23 & .152& Elementary particle theories && 61 & 0.21\% & 6
& 0& Networks \\ \hline

31 & 0.75\% & 22 & .242& Metal surface structure &&&&&&\\ \hline
\end{longtable}

}

\newpage

{\small
\begin{longtable}
{|p{0.3in}|p{0.3in}|p{0.5in}|p{2in}|p{0.1in}|p{0.3in}|p{0.3in}|p{0.5in}|p{2in}|}
\endhead
\caption{Highly-cited publications by decades.  The numerical label indicates
  the individual communities in Fig.~\ref{decades} and $N$ is number of
  publications in each community.  Each is also categorized by its Phys.\
  Rev.\
  classification and by its topic.}\label{tab-decade-1}\\
\hline \multicolumn{4}{|c|} {1927--36} &&  \multicolumn{4}{c|}{1937--46}\\ \hline
label & $N$ & class& topic && label & $N$ & class&topic\\ \hline
201 & 55 & PRA/E& Ionization/electron scattering && 301 & 52 & PRA/E& Radioactivity \\ \hline
202 & 104 &  PRA/E& Electron spectroscopy && 302 & 151 & PRC &Nuclear moments \\ \hline
203 & 64 &  PRB & X-Ray diffraction && 303 & 120 & PRC& Radioactivity/nuclear reactions \\ \hline
204 & 64 &  PRA/E& Infrared spectrum  && 304 & 121 & PRA/E&Nuclear physics/Cosmic rays \\ \hline
205 & 143 &  PRA/E& Atomic spectra && 305 & 90 & PRC&Gamma/beta rays \\ \hline
206 & 53 &  PRB& Photoelectric properties of metals && 306 & 133 & PRC& Neutron scattering \\ \hline
207 & 71 &  PRA/E& Ionization && 307 & 85 & PRA/E&Deuteron binding \\ \hline
208 & 79 &  PRA/E& Molecular spectra && 308 & 84 & PRD& Cosmic rays \\ \hline
209 & 57 &  PRB & Vacuum arc cathode && 309 & 54 & PRA/E& Mesons \\ \hline
210 & 106 &  PRA/E& Thermionic emission && 310 & 71 & PRA/E& Nuclear magnetic moments \\ \hline
211 & 117 &  PRA/E& X-ray spectra && && & \\ \hline
212 & 198 &  PRA/E& Cosmic ray/N \& P interactions &&& & & \\ \hline
213 & 65 &  PRA/E& Cosmic radiation && && & \\ \hline
214 & 51 &  PRB & Photoelectric effect &&& & & \\ \hline
215 & 133 &  PRA/E& Slow neutrons && && & \\ \hline
216 & 53 &  PRA/E& Spectrography && && & \\ \hline
\multicolumn{9}{c}~\\ 
\hline \multicolumn{4}{|c|} {1947--56} &&  \multicolumn{4}{c|}{1957--66}\\ 
\hline label & $N$ & class& topic && label & $N$ & class&topic\\ \hline

401 & 53 & PRC & Pion \& nucleon interactions && 501 & 121 & PRB &quantum magnetism \\ \hline
402 & 77 & PRB& Nuclear magnetic moments && 502 & 117 & PRD &Axial vector/weak interactions \\ \hline
403 & 68 & PRC & Nuclear levels/scattering && 503 & 52  & PRA/E &Optical beams \\ \hline
404 & 201 & PRC& Nuclear energy levels && 504 & 208 & PRB &Correlated electron systems \\ \hline
405 & 177 & PRB &Semiconductors && 505 & 69  & PRA/E &Atomic structure \\ \hline
406 & 73 & PRB & Mesons && 506 & 58  & PRB &Spin relaxation \\ \hline
407 & 140 & PRD &Quantum electrodynamics && 507 & 126 & PRC &Scattering \\ \hline
408 & 138 & PRD &Beta-ray spectra&& 508 & 227 & PRB &Many-body systems \\ \hline
409 & 72 & PRC &Nuclear reactions and levels && 509 & 81  & PRB &Bismuth \\ \hline
410 & 184 & PRC &Properties of nuclei && 510 & 62  & PRB &High energy scattering \\ \hline
411 & 112 & PRA/E &Precision QED tests && 511 & 140 & PRD &Parity nonconservation \\ \hline
412 & 83 & PRC &High energy scattering && 512 & 106 & PRD &Pion interactions \\ \hline
413 & 72 & PRA/E &Magnetic moments && 513 & 96  & PRA/E &Atomic collisions \\ \hline
414 & 110 & PRB &Application of magnetism && 514 & 254 & PRD &Symmetries of elementary particles \\ \hline
415 & 167 & PRC &QED/Nuclear reactions/scattering && 515 & 111 & PRB &Energy bands \\ \hline
416 & 89 & PRB &Liquid helium && 516 & 161 & PRB &Superconductivity \\ \hline
417 & 109 & PRD &Cosmic rays/nuclear scattering && 517 & 126 & PRC &Interaction of radiation with matter \\ \hline
\multicolumn{9}{c}~\\ 
\hline \multicolumn{4}{|c|} {1967--76} &&  \multicolumn{4}{c|}{1977--86}\\ 
\hline label & $N$ & class& topic && label & $N$ & class&topic\\ \hline

601 & 135 & PRD &Weak interactions && 701 & 256 &  PRB &Metal compounds \\ \hline
602 & 90  &  PRD &High-energy collisions && 702 & 85  &  PRB &Soliton in polyacetylene \\ \hline
603 & 77  &  PRD &Strong interaction && 703 & 132 &  PRB &Conductance fluctuations/scaling \\ \hline
604 & 132 &  PRB &Dielectric properties of complex materials && 704 & 136 &  PRB &$2d$
electron gas \\ \hline
605 & 234 &  PRB &Metals && 705 & 89  &  PRA/E &Diffusion limited aggregation \\ \hline
606 & 51  &  PRB &Two-dimensional systems && 706 & 163 &  PRB &Correlated electron systems \\ \hline
607 & 52  &  PRD &Quarks && 707 & 53  &  PRB &Semiconductor superlattices \\ \hline
608 & 98  &  PRD &Scattering theory && 708 & 70  &  PRD &Cosmology \\ \hline
609 & 80  &  PRB &TTF-TCNQ && 709 & 67  &  PRD &Particle physics \\ \hline
610 & 89  &  PRA/E &Self-induced transparency && 710 & 194 &  PRD &Relativistic collisions/cosmology \\ \hline
611 & 237 &  PRB &Scaling theory && && & \\ \hline
612 & 105 &  PRB &Alloys &&& & & \\ \hline 
613 & 89  &  PRA/E &Atomic structure &&& & & \\ \hline
\end{longtable}
}

\newpage
{\small
\begin{longtable}
{|p{0.3in}|p{0.3in}|p{0.5in}|p{2in}|p{0.1in}|p{0.3in}|p{0.3in}|p{0.5in}|p{2in}|}
\endhead
\caption{Continuation of table~\ref{tab-decade-1}.}\label{tab-decade-4}\\
\hline \multicolumn{4}{|c|} {1987--96} &&  \multicolumn{4}{c|}{1997--2006}\\ \hline
label & $N$ & class& topic && label & $N$ & class&topic\\
\hline 801 & 407 & PRB &High-Tc oxides && 901 & 75  & PRA/E &B.E.Condensation in optical lattices \\ \hline
802 & 112 & PRB &Pseudopotentials && 902 & 83  & PRB &Manganites \\ \hline
803 & 96  & PRA/E &B.E.Condensation in atomic gas && 903 & 57  & PRB &Electron transport in exotic systems \\ \hline
804 & 56  & PRB &Fractional quantum Hall effect && 904 & 70  & PRB &Quantum magnetism/quantum dots \\ \hline
805 & 51  & PRB &Photonics && 905 & 255 & PRA/E &Vortex formation in BEC \\ \hline
806 & 118 & PRB &RG and Monte Carlo simulation && 906 & 52  & PRB &Spin-Hall effect \\ \hline
807 & 73  & PRB &Giant magnetoresistance && 907 & 52  & PRC &High-energy nuclear collision \\ \hline
808 & 75  & PRB &Flux lattice melting and High-Tc && 908 & 214 & PRD &Elemental particle theory \\ \hline
809 & 68  & PRB &Quantum dots && 909 & 184 & PRB & High-Tc \\ \hline 
810 & 119 & PRB &SOC/pinning in superconductors && 910 & 113 & PRD &Cosmology \\ \hline
811 & 58  & PRB &Vortex-glass superconductivity && 911 & 256 & PRA/E &Quantum information \\ \hline
812 & 84  & PRB &Semiconductor epitaxial growth && && & \\ \hline
813 & 163 & PRA/E &Quantum computation/teleportation &&&  & & \\ \hline
814 & 89  & PRB &Conductance fluctuations &&& & & \\ \hline
815 & 76  & PRD &Particle physics && && & \\ \hline
\end{longtable}
}

\end{document}